\documentclass[preprint,prl,aps,tightenlines,floatfix,showpacs]{revtex4}

\usepackage{graphics}
\usepackage{bm}
\usepackage{amsmath}
\usepackage{amssymb}
\begin{document}

  \title{Simple function form for n+${}^{208}$Pb total cross section
    between 5 and 600 MeV}
  \author{P. K. Deb}
  \email{pdeb@mps.ohio-state.edu}
  \affiliation{Department of Physics, The Ohio State University,
    Columbus, OH 43210, U.S.A.} 
  \author{K. Amos}
  \email{amos@physics.unimelb.edu.au}
  \affiliation{School of Physics, The University of Melbourne,
    Victoria 3010, Australia}
  \author{S. Karataglidis}
  \email{kara@physics.unimelb.edu.au}
  \affiliation{School of Physics, The University of Melbourne,
    Victoria 3010, Australia} 

  \date{\today}
  \begin{abstract}
    The  total cross  section for  neutron scattering  from $^{208}$Pb
    with energies between 5 and  600 MeV has been analyzed extending a
    previously defined simple function of three parameters to reveal a
    Ramsauer-like  effect  throughout  the  whole energy  range.  This
    effect  can be  parametrized in  a simple  way so  that it  may be
    anticipated that the complete function prescription will apply for
    total cross sections from other nuclei.
  \end{abstract}
  \maketitle

  Total cross sections  from the scattering of nucleons  by nuclei and
  for energies to 600 MeV or  more, are required in a number of fields
  of study in basic science as  well as many of applied nature. Often,
  those  cross  sections have  been  evaluated using  phenomenological
  optical  potentials and much  effort has  gone into  defining global
  sets  of  parameter values  for  optical  potentials  with which  to
  estimate cross sections as yet unmeasured. In a recent study, Koning
  and  Delaroche~\cite{Ko03} gave  a detailed  specification  of such.
  However, it would  be useful and convenient if  total cross sections
  could  be well  approximated  by  a simple  function  form. We  show
  herein,  in the  case of  neutron scattering  from  $^{208}$Pb, that
  there is a simple three parameter  function form one can use to form
  estimates without  recourse to optical  potential calculations. With
  data from  other nuclei  having so similar  an appearance,  the same
  form can be used for any target. Further, the required values of the
  three   parameters   of  that   function   form,  themselves   trend
  sufficiently smoothly  with energy that they may  be interpolated to
  estimate  any cross  section value  at energies  that have  not been
  measured.

  We  also investigate  whether a  simple functional  form  for energy
  variation of the parameters themselves may exist. But like the cross
  section at relatively low  incident energies, one parameter required
  to fit data exhibits a noticeably large scale structure. With actual
  data, that  effect, defined as the Ramsauer  effect, varies smoothly
  with  target mass  so that  in the  past it  has been  attributed to
  characteristics  of   the  nuclear  geometry   and  was  interpreted
  semi-classically~\cite{Pe62}  as  due  to the  interference  between
  parts of  the scattering wave  function passing through  the nuclear
  medium  with parts  that do  not. The  focus in  optical  model wave
  functions~\cite{Am66}  is another result  of such  interference. The
  variation of  the total  cross sections then  was formulated  as the
  Ramsauer  model. Herein we  use that  Ramsauer model  in conjunction
  with  the simple  three parameter  function form  and find  that the
  measured total cross-section data can be well fit.

  The  utility of the  Ramsauer model  has been  demonstrated recently
  with extended  versions used  to study the  isospin effect  noted in
  comparison of  total neutron cross sections from  select medium mass
  nuclei~\cite{An90} and  to interpret zero  angle (p,n) cross-section
  data~\cite{Ma97}.  Also,  by  using  Wick's  limit,  estimations  of
  neutron reaction cross sections were made~\cite{Di03}. This portends
  an  interesting  development  relevant   to  our  approach.  If  the
  functional  form scheme  gives  results satisfying  Wick's limit  to
  within a  few percent, then the approach~\cite{Di03}  to specify the
  total reaction cross section for the same wide span of energy may be
  used      with     some      confidence.      Inverse     scattering
  theory~\cite{Mi69,Ch89,Lu96} may  then be  used to define  a complex
  local  optical potential  that reproduces  those data  and  which is
  essentially free of any model prescription.

  While  the total  cross sections  and their  large  scale structures
  should be  the result of  a more sophisticated specification  of the
  optical potential describing  neutron-nucleus scattering, and indeed
  aspects of the Ramsauer effect have been elicited from a $g$-folding
  optical  potential~\cite{Am00,De04},  the  predicted  results  never
  quite   match  satisfactorily   the   observed  Ramsauer   structure
  especially at energies below 50~MeV.

  The total cross  sections for nucleon scattering from  nuclei can be
  expressed in terms of  partial wave scattering matrices specified at
  energies     $E\propto      k^2$,     namely     $S^{\pm}_l(k)     =
  \eta^{\pm}_l(k)e^{2i\Re\left[\delta^{\pm}_l(k)\right]    }$,   where
  $\delta^\pm_l(k)$  are  the (complex)  scattering  phase shifts  and
  $\eta^{\pm}_l(k)$  are   the  moduli   of  the  $S$   matrices.  The
  superscript designates $j = l\pm 1/2$. In terms of these quantities,
  the total cross section is
  \begin{align}
    \sigma_{\text{tot}}(E) & = \frac{2\pi}{k^2} \sum_l
    \sigma^{(l)}_{\text{tot}}(E) \nonumber \\
    & = \frac{2\pi}{k^2} \sum_l \left[ \left( l + 1 \right) \left\{1 -
      \eta^+_l(k) \cos\left\{ 2\Re\left[ \delta^+_l(k) \right]
      \right\} \right\} \right. \nonumber \\
      & \; \; \; + \left. l \left\{ 1 - \eta^-_l(k) \cos\left\{
    2\Re\left[ \delta^-_l(k) \right] \right\} \right\} \right] .
    \label{SumTOT}
  \end{align}
  There are equivalent forms for  the total reaction and total elastic
  cross    sections   and    a   study    of   such    cross   section
  data~\cite{Am02,De03}  established  that  the  partial  total  cross
  sections   may   be   described    by   a   simple   function   form
  \begin{align}
    \sigma^{(l)}_{\text{tot}}(E) \equiv \sigma^{(l)}_{\text{th}}(E) &
    = \left( 2l + 1 \right) \left[ 1 + e^{\frac{( l - l_0 )}{a}}
      \right]^{-1} \nonumber \\
    & + \epsilon \left( 2l_0 + 1 \right) e^{\frac{( l - l_0 )}{a}}
    \left[ 1 + e^{\frac{( l - l_0 )}{a}} \right]^{-2} \, .
    \label{Fnfor}
  \end{align}
  As with our previous studies,  this form for the total cross section
  is suggested by the values of the partial total cross sections found
  from   energy-dependent,  optical   potentials   generated  from   a
  $g$-folding   formalism~\cite{Am00}.   With   that  form   excellent
  reproduction of  the proton total  reaction cross sections  for many
  targets and over a wide  range of energies were found with parameter
  values that  varied smoothly with energy  and mass. For  the case of
  scattering  from   ${}^{208}$Pb,  Skyrme-Hartree-Fock  model  (SKM*)
  densities~\cite{Br00} have been used to form the $g$-folding optical
  potentials to  give the initiating parameter  values. That structure
  when  used to  analyze  proton and  neutron scattering  differential
  cross   sections  at   65   and  200   MeV   gave  quite   excellent
  results~\cite{Ka02}.  Indeed  those   analyses  were  able  to  show
  selectivity for  that SKM*  model of structure  and for  the neutron
  skin  thickness of  0.17 fm  that it  proposed. That  same structure
  model has  been used in forming $g$-folding  optical potentials from
  which our initial guess at  the structure of the partial total cross
  section values at all energies to 600 MeV were obtained.

\begin{figure}
    \scalebox{0.7}{\includegraphics*{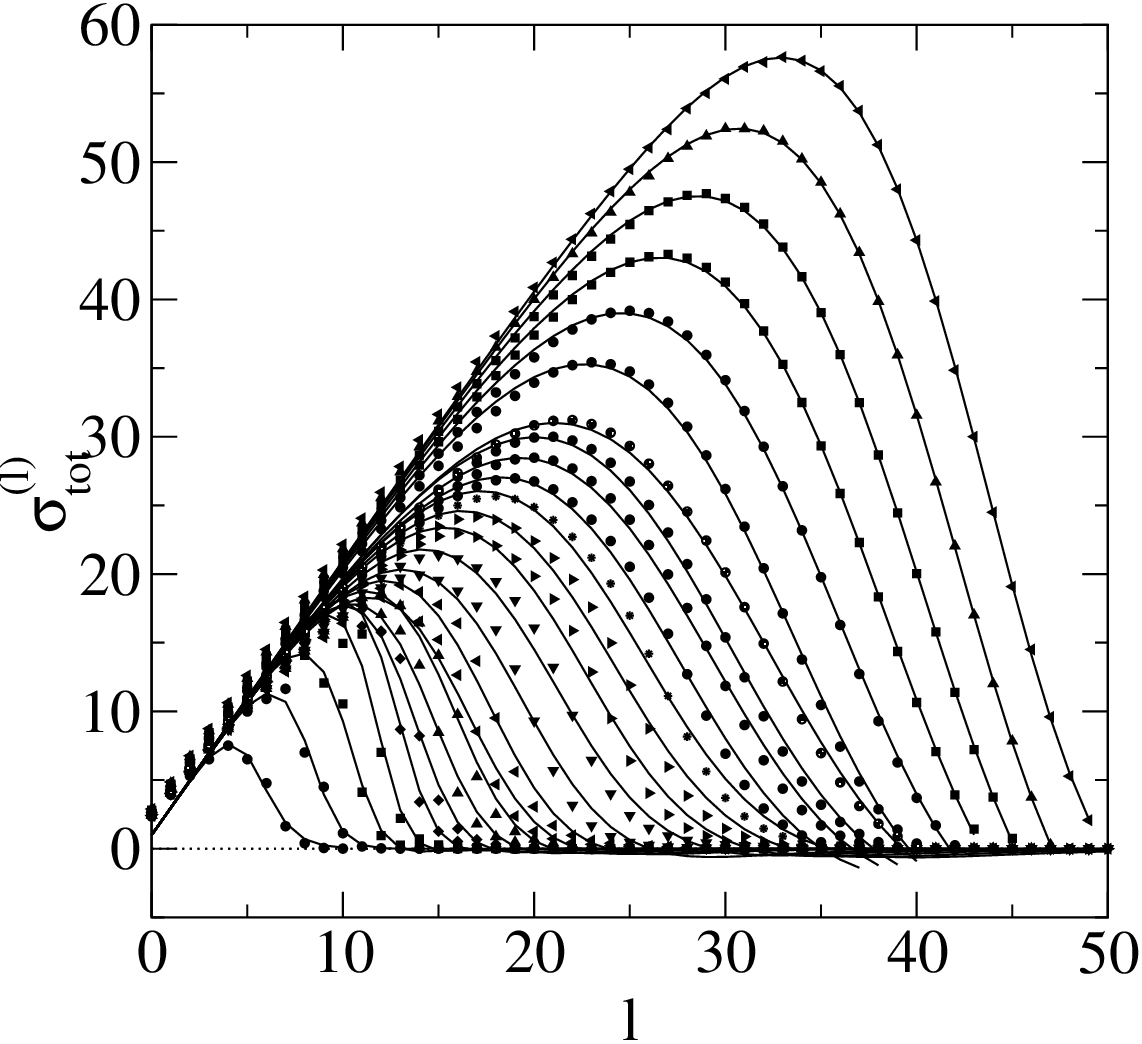}}
    \caption{\label{Fig1-DAK}  The partial  total  cross sections  for
      scattering of  neutrons from $^{208}$Pb for  energies between 10
      and  600 MeV.  The largest  energy  has the  broadest spread  of
      values.  The  'data'  were  obtained  from  $g$-folding  optical
      potential calculations.}
  \end{figure}

\begin{figure}
    \scalebox{0.7}{\includegraphics*{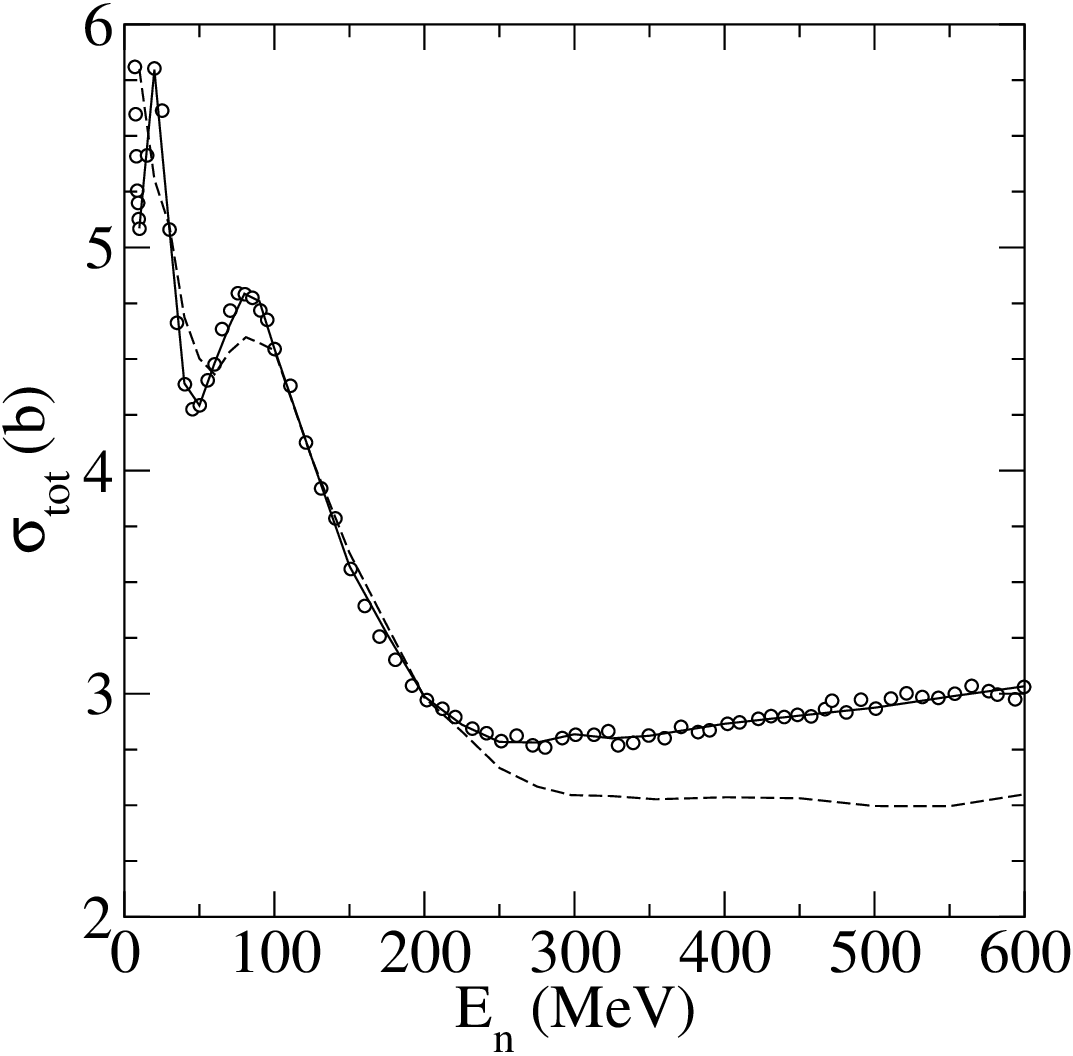}}
    \caption{\label{Fig2-DAK}
      Total cross sections for n-$^{208}$Pb scattering. The curves are
      as defined in the text.}
  \end{figure}

  The partial total cross  sections from $g$-folding optical potential
  calculations  for neutron  scattering from  $^{208}$Pb are  shown in
  Fig.~\ref{Fig1-DAK}.  Those values  are  shown as  diverse open  and
  closed symbols  in Fig.~\ref{Fig1-DAK} and we take  them as ``data''
  against which  to find the first  guess for the  energy variation of
  the three parameters in  Eq.~(\ref{Fnfor}). Each curve shown in that
  figure is  the result  of a search  for the  best fit values  of the
  three  parameters, $l_0$,  $a$,  and $\epsilon$.  From  the sets  of
  values that result from that fitting process, the two parameters $a$
  and  $\epsilon$  can  themselves   be  expressed  by  the  parabolic
  functions
  \begin{align}
    a & = \phantom{-}1.29 + 0.00250 E - 1.76 \times 10^{-6} E^2 \; ,
    \nonumber \\
    \epsilon & = -1.47 - 0.00234 E + 4.16 \times 10^{-6} E^2 \; ,
    \label{Eps}
  \end{align}
  With $a$  and $\epsilon$  so fixed, we  then adjusted the  values of
  $l_0$  in   each  case  so   that  actual  measured   neutron  total
  cross-section    data     were    fit    using    Eq.~(\ref{Fnfor}).
  The quality of  fits to measured total cross-section  data are shown
  in Fig.~\ref{Fig2-DAK}.
  Using the  SKM* model structure, the  $g$-folding optical potentials
  gave  the  total  cross  sections  shown  by  the  dashed  curve  in
  Fig.~\ref{Fig2-DAK}. Clearly  there is a need to  improve this model
  for energies  at and above  pion threshold. Nonetheless, it  does do
  quite  well for  lower energies,  most notably  giving  a reasonable
  account of the Ramsauer resonance~\cite{Ko03} near 100 MeV. However,
  recall that these $g$-folding values  serve only to provide a set of
  partial  cross  sections to  define  an  initial  set of  the  three
  parameters  of the  function form.  With $a$  and $\epsilon$  set by
  Eq.~(\ref{Eps}),  adjustment of  $l_0(E)$ produces  the  solid curve
  shown in Fig.~\ref{Fig2-DAK}; an excellent reproduction of the data,
  as it  was designed  to do. There  are obvious oscillations  in that
  tabulation  of  values  of  $l_0(E)$  for  energies  below  100  MeV
  reflecting the Ramsauer  effects in the cross section  data. But for
  energies above 250 MeV, the  $l_0$ values are well approximated as a
  straight line.  Without retaining the excellent fit  to values below
  100 MeV, a simple representation  of the $l_0$ values that is useful
  is          the          energy         dependent          function,
  \begin{equation}
    l_0^{\text{th}}(E)  =  0.0384  E  +  16.28  -  11.22  \left[  1  -
    \frac{E}{132.9} \right] e^{-0.0164 E}\; .
    \label{fnofE}
  \end{equation}
  With $a$ and $\epsilon$  specified by Eq.~(\ref{Eps}), this function
  form leads to an average background total cross section upon which a
  regular oscillatory contribution from  the Ramsauer effect is found.
  Of  course,  phenomenological   optical  potentials  by  appropriate
  parameter  adjustments~\cite{Ko03} will  define  the relevant  cross
  sections well for  all energies. But if a  global (smooth) variation
  of those parameter  values is made then the  quality of reproduction
  deteriorates.

\begin{figure}
    \scalebox{0.7}{\includegraphics*{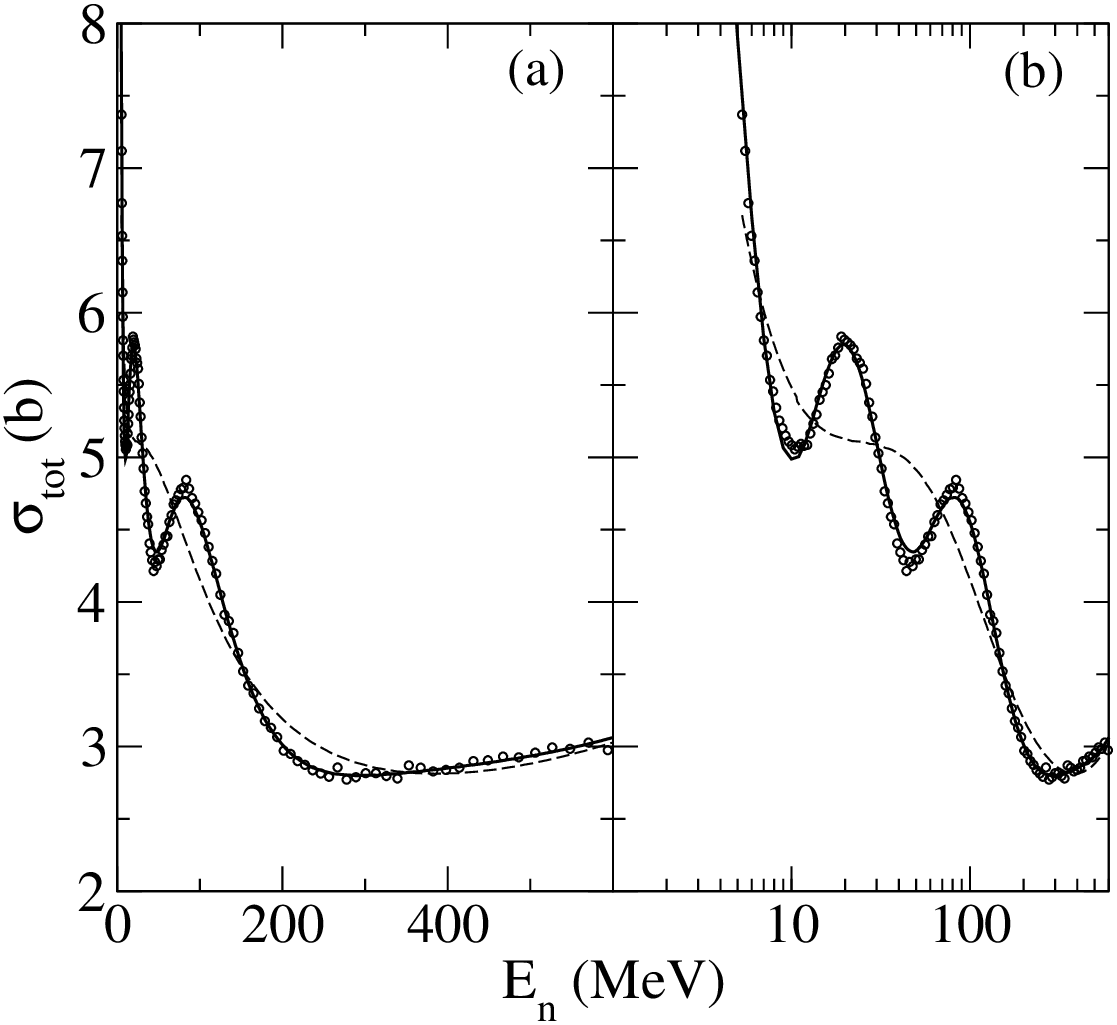}}
    \caption{\label{Fig3-DAK}  The  optimal  background cross  section
      with the data from  n-$^{208}$Pb scattering. The curves are as
      for Fig.~\ref{Fig2-DAK}.}
  \end{figure}

\begin{figure}
    \scalebox{0.7}{\includegraphics*{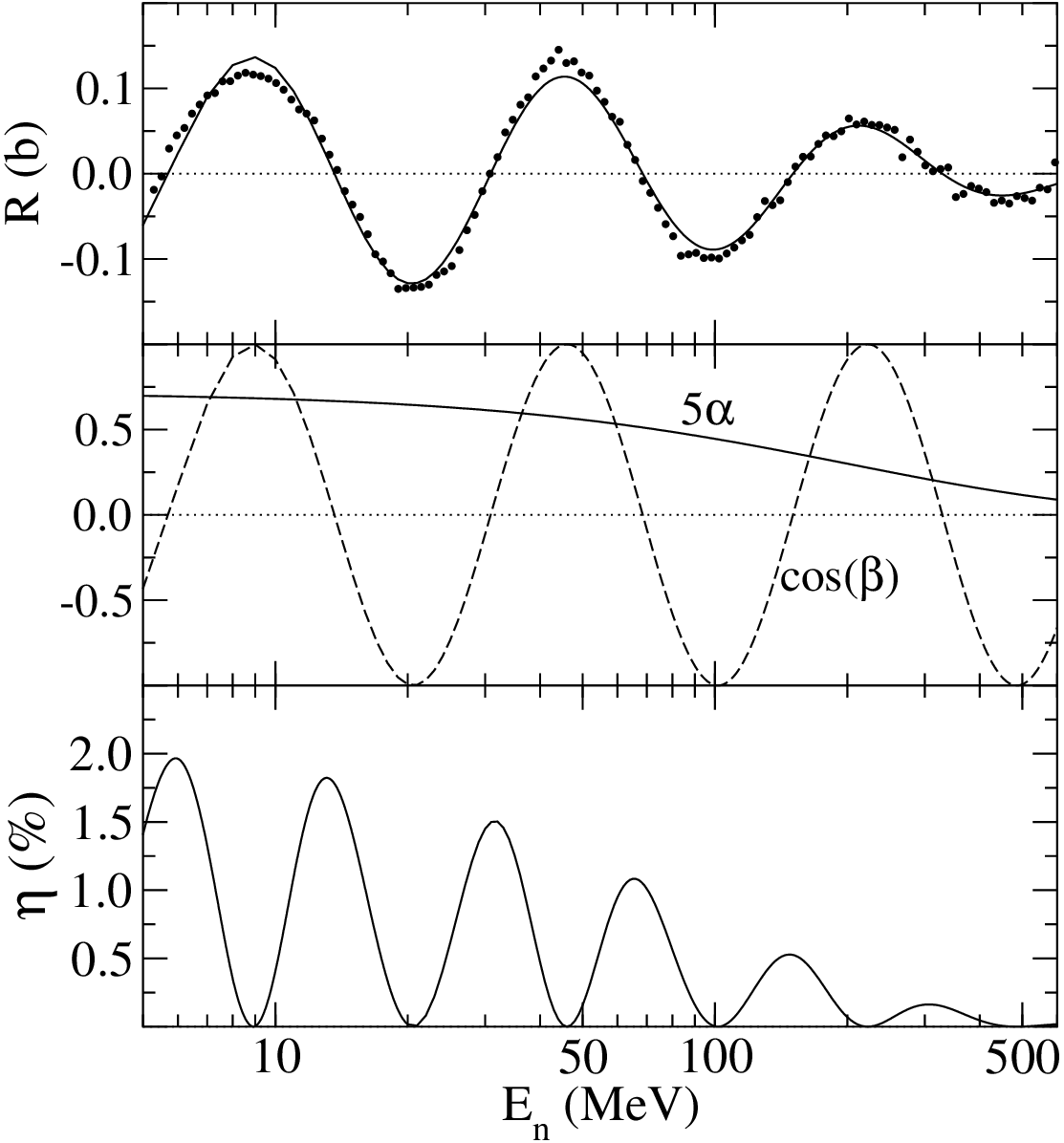}}
    \caption{\label{Fig4-DAK} The Ramsauer  corrections in data (top),
      parameters (middle), and  fractional deviation from Wick's limit
      (bottom) based upon the 3 parameter functional form model.}
  \end{figure}

  A Ramsauer-like  effect has been  included in past data  analyses to
  describe the large scale variations  of total cross sections from an
  otherwise    smooth    monotonic    background~\cite{Di03}.    Under
  approximation,  this   correction  is   a  coherent  scaling   of  a
  theoretical model (diffraction,  global optical, or functional form)
  of                the               background,               namely
  \begin{equation}  
    \sigma_{\text{tot}}(E) \equiv \sigma_{\text{th}}(E) \left[ 1 -
    \alpha(E) \cos\left( \beta(E) \right) \right] \; . 
  \end{equation}
  Dietrich {\it  et al.}~\cite{Di03} also linked this  by Wick's limit
  to  extract reaction  cross sections.  Wick's  limit is  that of  an
  inequality involving the zero  degree cross section and which arises
  from       the       optical       theorem,      namely,       since
  \begin{align}
    \Im{     \left[ f(0^\circ) \right]  }    &     =    \frac{k}{4\pi}
    \sigma_{\text{tot}}(E) \; , \nonumber \\
    \sigma( 0^\circ ) & = \left| f(0^\circ) \right|^2 \ge \left\{ \Im{
    \left[ f(0^\circ) \right] } \right\}^2 \ge \left[
      \frac{k}{4\pi}\sigma_{\text{tot}}(E) \right]^2 \; .
  \end{align}
  The  equality  specifies  the  cross  section  at  the  Wick  limit,
  $\sigma^W(0^\circ)$. Then as the Ramsauer model is based upon a form
  for          the           scattering          amplitude          of
  \begin{equation}
    f(\theta)  =   i  \frac{k}{4\pi}  \sigma_{\text{th}}(E)  \left(1  -
    \alpha(E) e^{i\beta(E)} \right)\; ,
  \end{equation}
  the zero degree cross section is 
  \begin{align}
    \sigma(0^\circ) & = \left[ \frac{k}{4\pi} \sigma_{\text{th}}(E)
    \right]^2 \left\{\left[1 - \alpha\cos(\beta)\right]^2 +
    \alpha^2\sin^2(\beta) \right\} \nonumber \\
    & = \sigma^W(0^\circ) \left\{1 + \frac{\alpha^2\sin^2(\beta)}
    {\left[ 1 - \alpha \cos(\beta) \right]^2}
    \right\}\; .
  \end{align}
  In this and  the next equation the energy  variable has been omitted
  for convenience. Thus the validity  of Wick's limit in this model is
  measured by the fractional deviation found for the zero degree cross
  section expectation
  \begin{equation}
    \eta(\%) = 100 \frac{ \left[ \sigma(0^\circ) - \sigma^W(0^\circ)
	\right] }{ \sigma^W(0^\circ) } = 100 \left[ \frac{ \alpha
	\sin(\beta) }{ 1 - \alpha \cos(\beta)  }\right]^2 \; .
	\label{fracdev}
  \end{equation}
  Results   for    n-$^{208}$Pb   scattering   are    displayed   in
  Figs.~\ref{Fig3-DAK} and \ref{Fig4-DAK}.  With the particular choice
  for the energy dependence of $l_0(E)$ as given in Eq.(~\ref{fnofE}),
  we note a Ramsauer effect  that continues to high energies but which
  has a  very regular  character. The total  cross sections  are shown
  linearly with energy in the top segment in Fig.~\ref{Fig3-DAK} while
  they  are  displayed on  a  logarithmic-linear  plot  in the  bottom
  segment. The  top segment illustrates the overall  smoothness of the
  data  variation and suggests  that the  functional form  results are
  very  good for  energies above  150  MeV. Indeed  adjustment of  the
  values in Eq.~(\ref{fnofE})  can give a much better  fit to the high
  energy data, but the chosen set  give a more regular behavior to the
  Ramsauer   effect   contributions.   The   logarithmic-linear   plot
  emphasizes the  low energy regime  and illustrates more  clearly the
  Ramsauer                                                      effect.
  The  bottom  panel also  reveals  that  the  Ramsauer effect  has  a
  sinusoidal variation with $\log_{10}(E)$  and that the wavelength is
  $\sim 0.7\log_{10}(E)$.

  In Fig.~\ref{Fig4-DAK}, details of the Ramsauer correction given the
  chosen background form are displayed.  All three graphs are shown in
  linear-logarithmic  form  to  emphasize   the  scale  in  which  the
  corrections are nearest to sinusoidal.  In the top panel we show the
  factor           $R(E)          =          \alpha(E)\cos(\beta(E))$.
  Clearly the  Ramsauer variation has  a regular behavior  with energy
  with the  wavelength indicated  above. Also in  this scale,  the two
  parameters, $\alpha(E)$, and  $\beta(E)$, vary smoothly with energy.
  The former is bounded by  $\pm 0.2$ and monotonically decreases with
  energy. Finally in the bottom panel we show the fractional deviation
  from Wick's limit as defined by Eq.~(\ref{fracdev}). Never does this
  analysis exceed a  2 \% violation of that limit  and so the approach
  of Dietrich \textit{et  al.}~\cite{De03} to extract neutron reaction
  cross  sections  from  the   total  cross  sections,  and  with  the
  reasonable  accuracy that those  authors require,  may be  used then
  with the functional form for the background cross section.

  Of course the  actual effect displayed depends upon  what is used as
  the  smooth background.  However,  the indicated  maxima and  minima
  should be a good first  order estimate. These results also show that
  the Ramsauer  effect is not necessarily constrained  to low energies
  as one  may have  assumed. Indeed save  for very low  energies where
  channel coupling  and discrete resonance features  in scattering are
  expected  to be  important,  the effect  has  been and  should be  a
  feature of  the most appropriate  optical potential~\cite{An90,Di03}
  and    which    should   be    complex,    energy   dependent    and
  non-local~\cite{Am00}.  Such  if formed  microscopically~\cite{Am00}
  require, at the very least, an appropriate complex medium and energy
  dependent  effective interaction  between the  incident  nucleon and
  each and every  bound target nucleon as well  as a significant large
  basis,  many nucleon  prescription of  that  target. That  is a  not
  insignificant endeavor  though for  $^{208}$Pb, just such  an effort
  lead  to  aspects of  the  Ramsauer  structure  in the  total  cross
  section~\cite{De04}  as  well as  showing  that  analyses of  proton
  scattering  data provide  a very  reasonable criteria  for  the skin
  thickness~\cite{Ka02}.

  We suggest a  simple function form for partial  total cross sections
  whose  sum,   when  scaled  by   the  Ramsauer  effect,   will  give
  neutron-$^{208}$Pb  total  cross  sections  for any  energy  without
  recourse to  phenomenological optical potential  parameter searches.
  That  basic  function form  also  reproduces  proton reaction  cross
  sections.  The parameters that  fit actual  data show  smooth trends
  with  both energy.  Our results  suggest that  the  Ramsauer effect,
  visually  obvious  for  energies  below  100 MeV,  persists  at  all
  energies.  Further with  our functional  form, Wick's  limit remains
  valid to  within 2\% and so may  be used to find  the total reaction
  cross   section  by   using  the   method  of   Dietrich  \textit{et
  al.}~\cite{Di03}. With  both data, inverse scattering  theory may be
  used  to specify an  optical potential  without recourse  to defined
  radial forms.  Finally, given the  similarity in the shape  of total
  cross section data from many nuclei, the method should be universal.

  This research  was supported by research grants  from the Australian
  Research Council and by  the National Science Foundation under Grant
  No. 0098645.

  \bibliography{PRLdraft}

\end{document}